\documentclass[twocolumn,english,aps,prl,noshowpacs,floatfix,superscriptaddress]{revtex4}
\usepackage[T1]{fontenc}
\usepackage[latin9]{inputenc}
\usepackage{amsmath}
\usepackage{amssymb}
\usepackage{graphicx}
\usepackage{float}
\usepackage{array}
\newcolumntype{C}[1]{>{\centering\let\newline\\\arraybackslash\hspace{0pt}}m{#1}}
\usepackage{multirow}

\makeatletter
\@ifundefined{textcolor}{}
{%
 \definecolor{BLACK}{gray}{0}
 \definecolor{WHITE}{gray}{1}
 \definecolor{RED}{rgb}{1,0,0}
 \definecolor{GREEN}{rgb}{0,1,0}
 \definecolor{BLUE}{rgb}{0,0,1}
 \definecolor{CYAN}{cmyk}{1,0,0,0}
 \definecolor{MAGENTA}{cmyk}{0,1,0,0}
 \definecolor{YELLOW}{cmyk}{0,0,1,0}
}

\newcommand*{\balancecolsandclearpage}{%
  \close@column@grid
  \clearpage
  \twocolumngrid
}

\@ifundefined{definecolor}
 {\usepackage{color}}{}
\usepackage{siunitx}

\makeatother

\usepackage{babel}
\begin{document}
\flushbottom

\title{Using Shor's algorithm on near term Quantum computers: a reduced version}

\author{Martina Rossi}
\affiliation{Data Reply s.r.l., Corso Francia, 110, 10143 Turin, ITALY}

\author{Luca Asproni}
\affiliation{Data Reply s.r.l., Corso Francia, 110, 10143 Turin, ITALY}

\author{Davide~Caputo}
\affiliation{Data Reply s.r.l., Corso Francia, 110, 10143 Turin, ITALY}
\affiliation{Department of Mathematics and Physics - University of Salento, Via Arnesano, 73100 Lecce, ITALY}

\author{Stefano Rossi}
\affiliation{Data Reply s.r.l., Corso Francia, 110, 10143 Turin, ITALY}

\author{Alice Cusinato}
\affiliation{Data Reply s.r.l., Corso Francia, 110, 10143 Turin, ITALY}

\author{Remo Marini}
\affiliation{Generali Italia S.p.A., Via Marocchesa, 14, 31021, Mogliano Veneto, ITALY}

\author{Andrea Agosti}
\affiliation{Generali Italia S.p.A., Via Marocchesa, 14, 31021, Mogliano Veneto, ITALY}

\author{Marco~Magagnini}
\affiliation{Data Reply s.r.l., Corso Francia, 110, 10143 Turin, ITALY}

\begin{abstract}
	\section*{Abstract}
	Considering its relevance in the field of cryptography, integer factorization is a prominent application where Quantum computers are expected to have a substantial impact. Thanks to Shor's algorithm this peculiar problem can be solved in polynomial time. However, both the number of qubits and applied gates detrimentally affect the ability to run a particular quantum circuit on the near term Quantum hardware. In this work, we help addressing both these problems by introducing a reduced version of Shor's algorithm that proposes a step forward in increasing the range of numbers that can be factorized on noisy Quantum devices. The implementation presented in this work is general and does not use any assumptions on the number to factor. In particular, we have found noteworthy results in most cases, often being able to factor the given number with only one iteration of the proposed algorithm. Finally, comparing the original quantum algorithm with our version on simulator, the outcomes are identical for some of the numbers considered.
\end{abstract}

%\date{\today}
\maketitle

\section{introduction}
The security of many asymmetric cryptographic systems relies on the difficulty of some mathematical problems \cite{suo_quantum_2020}; this entails an exponential time for finding the private key. Among them, integer factorization represents the core of the RSA cryptosystem \cite{rivest_method_1978}. More specifically, RSA encryption and decryption mechanisms apply modular exponentiation using three values: $e$, $n$, and $d$. The couple $(e, n)$ represents the public key, while $d$ is the private key. $n$ is computed as the product of two large prime numbers, $p$ and $q$, while $d$ is derived from these factors. Therefore, if Bob wants to send an encrypted message to Alice, he must use her public key. Alice will obtain the plain message applying the same modular exponentiation using $d$. As a consequence, the security of the private key $d$ depends on the difficulty of finding $p$ and $q$ from $n$. However, the state-of-the-art classical fastest approach to solve this problem is the Number Field Sieve sub-exponential algorithm \cite{buhler_factoring_1993, crandall_prime_2005}, which requires a number of operations that exponentially increases with the dimension of the number to be factored. \\\\
As a consequence, the development of Shor's algorithm \cite{shor_polynomial-time_1997} in the Quantum Computing field proves particularly interesting. Indeed, this approach solves the integer factorization problem in polynomial time. Therefore, Shor's algorithm is the focus of many studies that aim to implement it on Quantum Processing Units (QPUs). Most of them propose an ad-hoc quantum circuit for $N=15$ \cite{vandersypen_experimental_2001, lanyon_experimental_2007, lu_demonstration_2007, politi_shors_2009, monz_realization_2016, lucero_computing_2012} executed on different technologies such as photonic systems or superconducting qubits. Qubit recycling, which consists in re-setting qubits to $|0\rangle$ after using them \cite{anikeeva_recycling_2021}, is applied in \cite{martin-lopez_experimental_2012} to factor $N=21$, while a simplified version of the algorithm is proposed in \cite{geller_factoring_2013} to factor $51$ and $85$ as examples of products of Fermat primes. A compiled version of Shor's algorithm is described in \cite{smolin_pretending_2013}, where specific values of the parameter $a$ are considered, since they always produce a trivial period. \\\\
Implementing the general Shor's algorithm on quantum hardware is challenging both because of the number of required qubits and the circuit depth. This second aspect can be particularly problematic because the quantum decoherence limits the number of gates that can be applied on each qubit. Moreover, the deeper the circuit, the greater is the introduced error. This explains the large presence of specific solutions for certain values of $N$ or $a$ in the literature. In this context, we propose an approximated version of Shor's algorithm that helps addressing the qubits and depth problems previously described. This reduces the hardware requirements in terms of supported depth; therefore, it is possible to factor bigger numbers on near term QPUs, compared to the classical version of Shor's algorithm. Finally, we did not consider specific values for $N$ or the parameter $a$.\\\\
Shor's algorithm is firstly presented in Section \ref{sec:algorithm} and its implementation using $2n+3$ qubits \cite{beauregard_circuit_2003} is described in Section \ref{sec:impl}. Section \ref{sec:approach} introduces our reduced version; this is firstly tested on simulator, to compare its performance with the original circuit. Moreover, performance on QPU was studied as well. The overall results are collected in Section \ref{sec:exper}, while these are thoroughly discussed in Section \ref{sec:disc}. Finally, conclusions and future works are presented in Section \ref{sec:concl}.

\section{Shor's algorithm} \label{sec:algorithm}
The integer factorization problem can be solved applying the algorithm proposed by Shor in 1994 \cite{shor_polynomial-time_1997}. This procedure combines classical and quantum computation; it consists in a classical pre-processing, a quantum circuit and a classical post-processing of the output of the quantum circuit. More specifically, the factorization problem is translated in an order-finding procedure. Given the function $f(x) = a^{x}mod N$, where $\{a\in R | 1<a<N\}$, the order of $f(x)$ is defined as the smallest positive integer $r$ such that $a^{r}mod N = 1$. The order-finding problem is solved applying the Quantum Phase Estimation (QPE) to the unitary operator $U_{a}|x\rangle = |(a\cdot x)mod N\rangle$ \cite{nielsen_quantum_2010}. Therefore, this is the quantum core of the overall algorithm.\\\\
Considering an odd integer $N$, which is not a prime power, Shor's algorithm identifies a factor with probability at least $1-\frac{1}{2^{k-1}}$, where $k$ represents the total of prime factors of $N$ \cite{shor_polynomial-time_1997}. The detailed procedure is described in the following \cite{nielsen_quantum_2010}: \\\\

\paragraph{Shor's algorithm}
\begin{enumerate}
  \item Preliminary checks on $N$: verify that $N$ is not even or a prime power; return the trivial factor in this case.
  \item Generate a random integer $a$ such that $1<a<N$ and compute $d = gcd(N, a)$. If $d > 1$, it is already a factor of $N$; thus, return $d$. \label{restart_point}
  \item Execute the order-finding procedure which applies the Quantum Phase Estimation circuit and the continued fraction algorithm to estimate $r$.
  \item If $r$ is even, compute $d_{1} = gcd(a^{r/2}-1, N)$ and $d_{2} = gcd(a^{r/2}+1, N)$. Verify that $d_{1}$ and $d_{2}$ are non-trivial factors. Otherwise, restart the algorithm from \ref{restart_point}.
\end{enumerate}

\begin{figure*}[t!]
	\centering
	\includegraphics[width=0.9\textwidth]{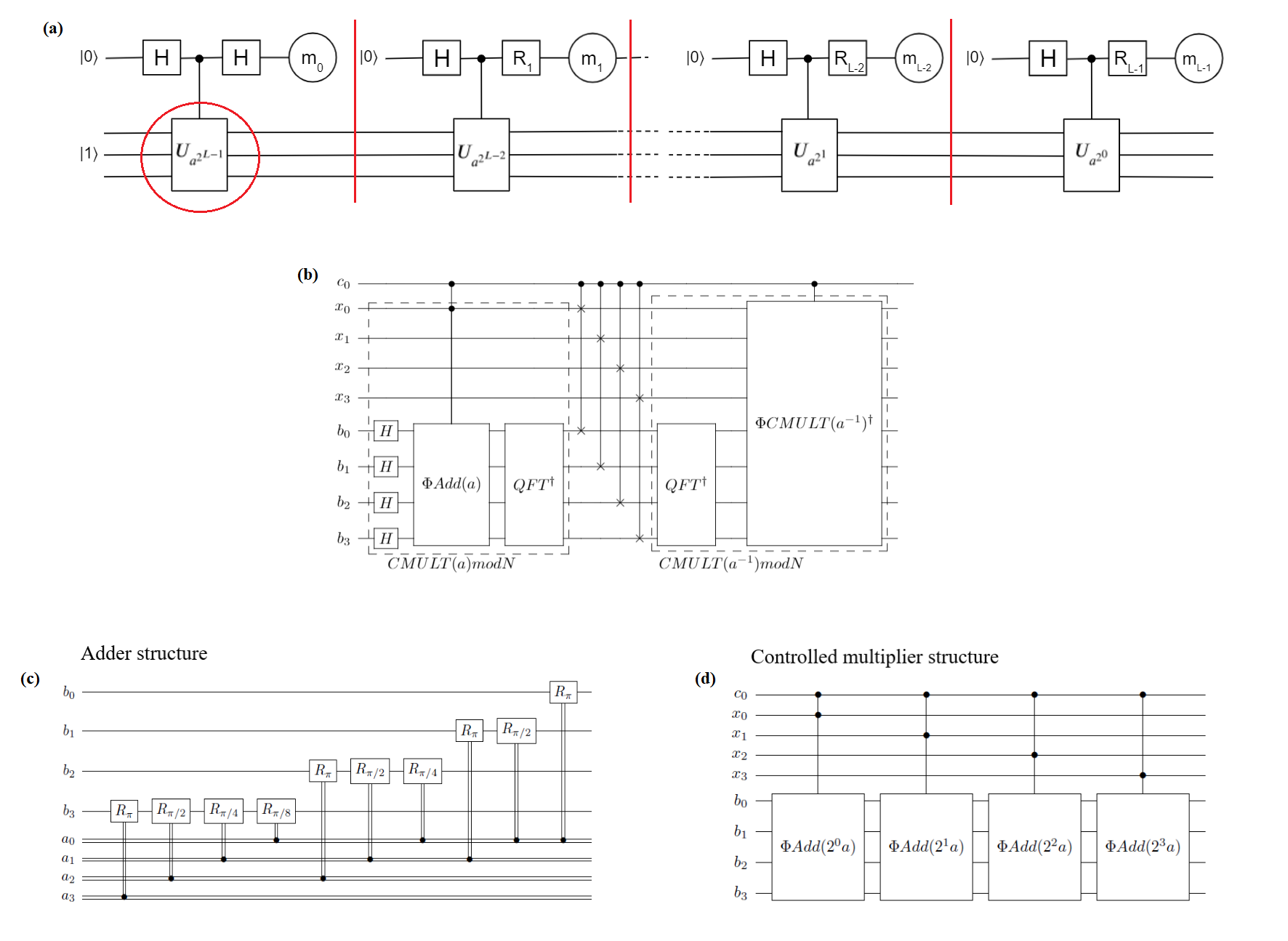}
	\caption{\textbf{Structure of the quantum circuit that implements the proposed version of Shor's algorithm.} \textbf{(a)} quantum circuit implementation for Shor's algorithm using only one control qubit. The control qubit is repeatedly initialized and measured to obtain each bit of the estimated phase. $R_i$ gates rotate the control qubit to apply a partial Inverse Fourier Transform, depending on the outcome of the previously measured values. More specifically,  $R_{j} = \begin{pmatrix}
        1 & 0\\
        0 & e^{i\theta_{j}}
    \end{pmatrix}$ and $\theta_{j} = -2\pi \sum_{k=0}^{j}\frac{m_{k}}{2^{j-k+1}}$ \cite{parker_efficient_2000, beauregard_circuit_2003}.
	The original quantum circuit is illustrated with graphical description of how it is subdivided in sub-circuits. Each unitary block has three macro-components: a controlled multiplier, a controlled swap and an inverse controlled multiplier which uses the modular inverse of $a$. The input qubits of the unitary block are initialized to $|1\rangle$. This unitary block is approximated with the circuit detailed in \textbf{(b)}. The corresponding graph highlights the macro components. The quantum adder gate $\Phi Add(a)$ \cite{draper_addition_2000} is described in \textbf{(c)}, while the controlled multiplier $\Phi CMULT(a)$ \cite{draper_addition_2000, vedral_quantum_1996} is illustrated in \textbf{(d)}. These last three circuits consider $n=4$ for greater clarity; the architecture can be easily generalized to any value of $n$.
    }
	\label{fig:2}
\end{figure*}

\section{original implementation} \label{sec:impl}
Considering the overall Shor's algorithm, the implementation of the QPE circuit represents the most critical part. More specifically, the QPE classical procedure uses two quantum registers to estimate the phase of an eigenvalue of a unitary operator, applied on the related eigenstate \cite{kitaev_quantum_1995}. The first register will contain the measured phase at the end of the execution; therefore, the precision required by the phase estimation determines the number of qubits in the register \cite{nielsen_quantum_2010, rieffel_quantum_2011, cleve_quantum_1998}. Typically, this register uses $2n$ qubits, where $n = \lfloor log_{2}(N)\rfloor + 1$ and $N$ is the number to be factored. Conversely, the second register contains the eigenstate used for the phase estimation; for the specific case of the order-finding problem, a convenient eigenstate is $|1\rangle$. \\\\
The first register is initialized in superposition, using Hadamard gates; then, these qubits are used to control different unitary operators that realize the modular exponentiation required by the order-finding procedure. Thanks to the phase kickback, the phase of the unitary operator is in the first register at the end of this step. Then, the Inverse Quantum Fourier Transform is applied to obtain the phase value in the computational basis, before measuring it.\\\\
Considering the classical implementation, each qubit in the first register controls only one unitary operator; therefore, a first optimization consists in using only one control qubit \cite{beauregard_circuit_2003, mosca_hidden_1999}. This qubit is iteratively initialized, then controls one unitary operator, undergoes a partial Inverse Quantum Fourier Transform and it is measured. After the measurement, the qubit is re-initialized to $|0\rangle$. This sequence of operations is repeated $2n$ times, to obtain the same precision on the phase measured, compared to the extensive approach. This optimized version of the order-finding circuit is illustrated in Figure \ref{fig:2}.a. \\\\
As it is described in \cite{beauregard_circuit_2003}, each unitary block is composed by three macro-components: a controlled-multiplier, a controlled swap and an inverse controlled-multiplier. Every multiplier is obtained as a sequence of adder modulo N. Moreover, there is a Quantum Fourier Transform (QFT) at the beginning of the block and an Inverse QFT at the end, since quantum adders operate in the Fourier basis. Finally, modular adders are realized manipulating a classical adder and using 2 ancilla qubits. Regarding these two last qubits, the former is used to expand the quantum register of the adder, to prevent overflow; the latter is necessary to obtain a modular adder from classical adders. It is clear that each unitary block is a complicated sub-circuit; therefore, $2n$ unitary blocks entail a considerable depth. \\\\
Recycling one qubit significantly reduces the overall number of qubits needed. More specifically, this circuit requires $2n+3$ qubits, where most of them are manipulated in the unitary block; indeed, only one qubit is used as control qubit. However, this circuit has the same depth of the more extensive version of the order-finding circuit. Consequently, the circuit in Figure \ref{fig:2}.a would face some challenges in running on real hardware, even if the number of qubits allowed it. Moreover, there might be an additional issue regarding the implementable gates. The partial Inverse QFT requires quantum gates controlled by classical bits whose values are the outcome of previous measurements; for example, these gates are not currently implementable on the IBM hardware. \\\\

\begin{figure*}[t!]
	\centering
	\includegraphics[width=0.98\textwidth]{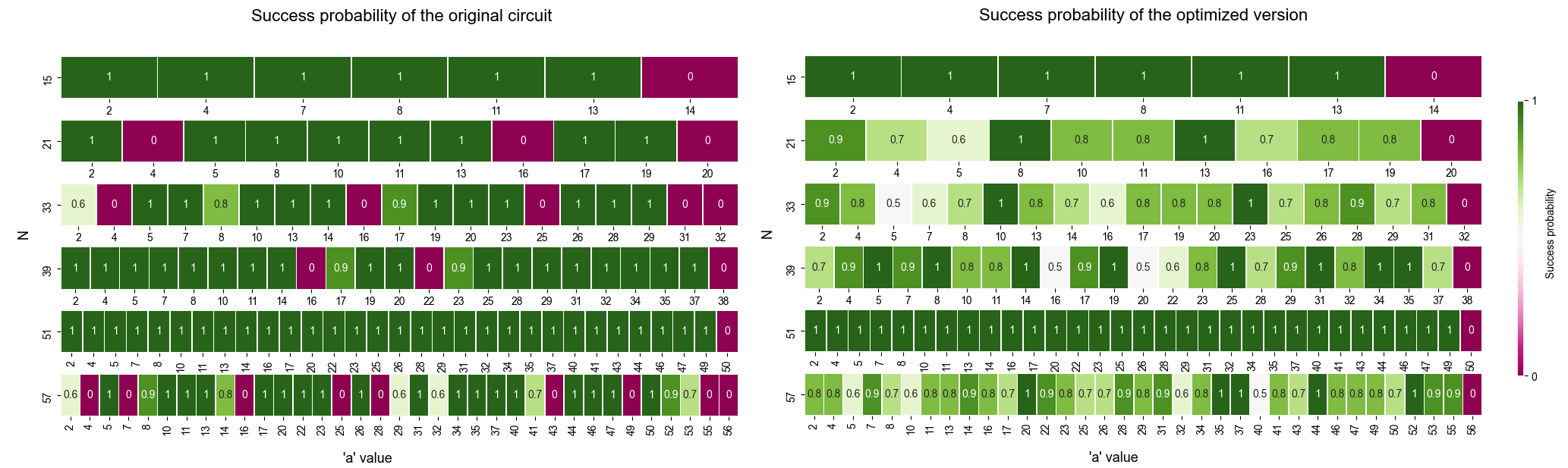}
	\caption{\textbf{Success probability heat maps comparing the performance of the proposed approach with the original circuit.}
		Each graph considers all the feasible values of 'a' for $N\in \{15,21,33,39,51,57\}$. The success probability is computed considering the frequency with which the corresponding circuit factorized N in one iteration of the algorithm, using the given 'a'. The study considers 10 repetitions for each couple $(N,a)$ and these are executed on the IBM '$ibmq\_qasm\_simulator$' simulator. }
	\label{fig:3}
\end{figure*}

\section{proposed approach} \label{sec:approach}
The goal of the proposed approach is to reduce the circuit depth, starting from the $2n+3$ order-finding circuit previously described. The first change consists in splitting the overall circuit into $2n$ sub-circuits. As it is illustrated in Figure \ref{fig:2}.a, each sub-circuit initializes the control qubit, applies one unitary block, executes the partial Inverse QFT and measures the control qubit. This first approximation aims to tackle both the depth issue and the use of potentially problematic gates, not always implementable in current hardware.\\\\
However, even these reduced circuits can be too deep and consequently we still have quantum dechoerence issues. This is mostly due to the operations in the unitary block implementation. Therefore, we thoroughly investigated the circuit composition for the unitary operator in order to identify which elements could be approximated without significantly affecting the accuracy of the outcome.  \\\\
As previously mentioned, each unitary block is composed by a controlled-multiplier, a controlled swap and an inverse controlled-multiplier. Figure \ref{fig:2}.b shows the structure of the approximated unitary operator, where the quantum register $x$ is initialized to $|1\rangle$, while the register $b$ is set to $|0\rangle$; we firstly focused on the controlled-multiplier. Because of this initialization, the initial QFT applied on $|b\rangle$ can be simplified in a sequence of Hadamard gates. Moreover, only one adder is applied on $|b\rangle$, since only one qubit of $|x\rangle$ is different from $0$. \\\\
We also found a simplified version of the inverse controlled-multiplier. The last QFT inside this block which is usually applied on $|b\rangle$ is removed, since this operation is not controlled by the control qubit and the circuit is split. Finally, all the adders modulo $N$ considered in this circuit are replaced by a simple adder. This last approximation entails a reduction of 2 qubits; resulting in a circuit that requires $2n+1$ qubits. The structure of the simple adder is described in Figure \ref{fig:2}.c. This last simplification might produce the most significant impact on the final outcome. However, the benefits of a shallower circuit outweigh the problems associated with a less accurate phase. 

\section{results}\label{sec:exper}
\begin{figure}[htbp]
	\centering
	\includegraphics[width=0.5\textwidth]{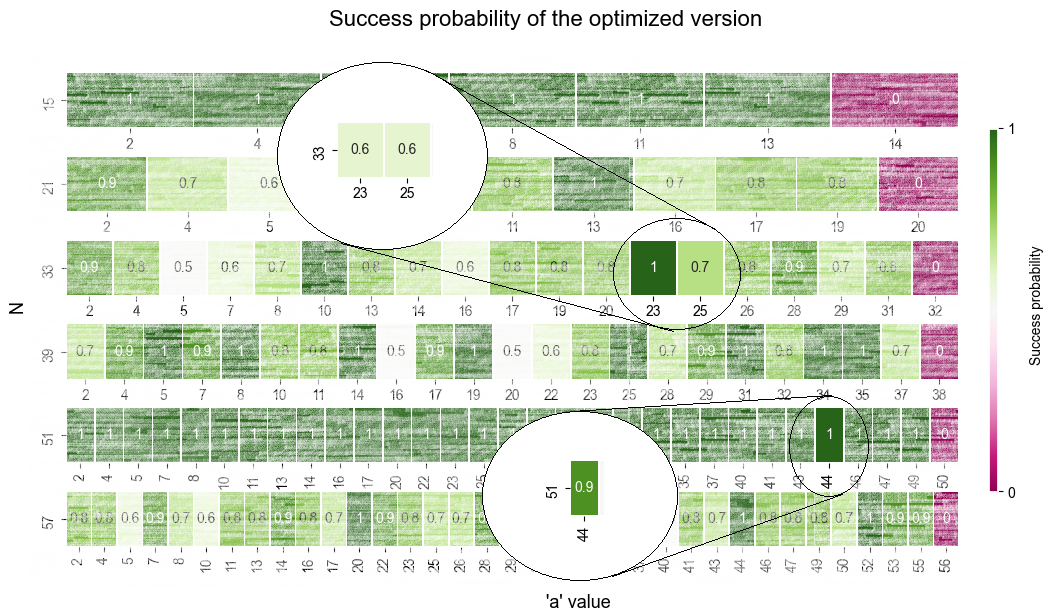}
	\caption{\textbf{Success probability heat map of the optimized version executed on QPU.} The proposed approach was tested on QPU for $(N = 51, a = 44)$, $(N = 33, a = 23)$,  and  $(N = 33, a = 25)$. This graph compares the performance of the proposed circuit on different devices, considering the same parameters.
		}
	\label{fig:heatmap_qpu}
\end{figure}

The performance of the proposed circuit was firstly compared to the behaviour of the original $2n+3$ circuit on the IBM simulator $ibmq\_qasm\_simulator$. This testing environment was necessary because the original circuit can be executed only on a simulator; therefore, a complete study of our circuit was realized on the same setting, for a fair comparison. We defined the success probability as the frequency of times (out of 10 experiments) where we could factor a given number $N$ in a single iteration, i.e. it factors immediately with the given $a$. We tracked the results of each circuit for multiple values of $N$ and the parameter $a$. It is worth noting that the proposed approach is always able to factor in less than 10 tries. This information in collected in Figure \ref{fig:3}.\\\\
Overall, our approach has a high success probability on simulator. Furthermore, we tested the circuit on a real hardware as well. We executed it on the IBM QPU $ibmq\_16\_melbourne$. Only few combination of $N$ and $a$ were tested, to verify the effect of a noisy hardware on our implementation. The outcomes are described in Figure \ref{fig:heatmap_qpu}.

\begin{figure*}%[htbp]
	\centering
	\includegraphics[width=0.9\textwidth]{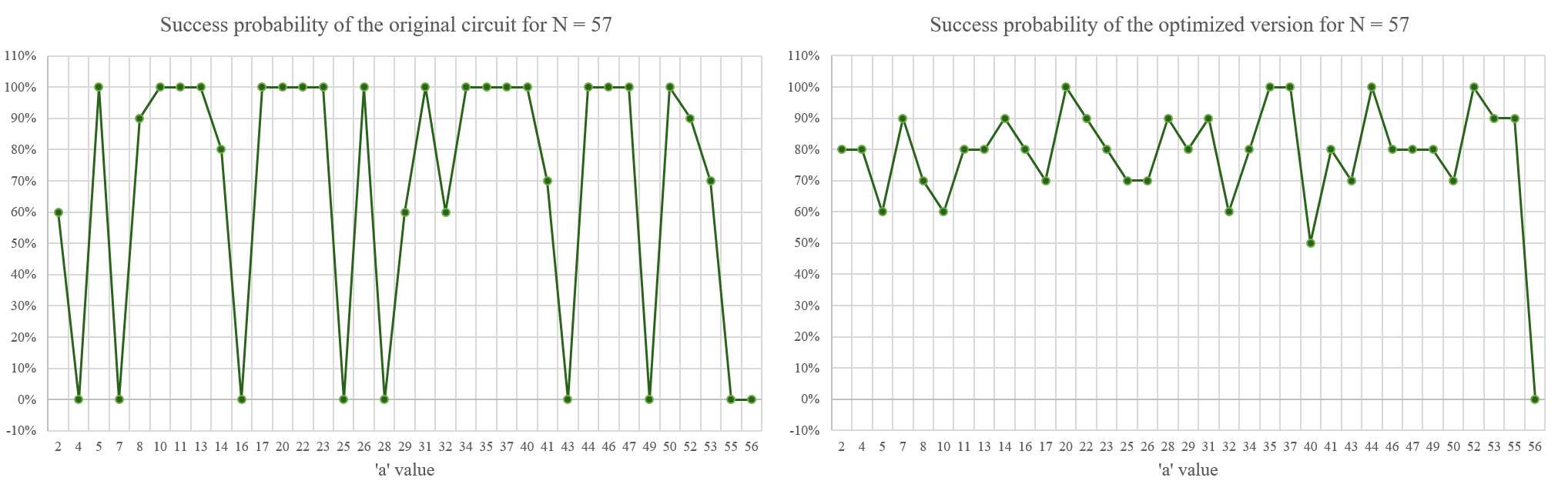}
	\caption{\textbf{Cross section of the success probability for a given $N$, considering all the feasible values of 'a'. } The original circuit has a highly fluctuating behaviour with some values of 'a' preventing factorization. Conversely, the proposed version is characterized by a less varied tendency and a success probability greater than 0 for most of the 'a' values which correspond to failures in the original circuit.
		}
	\label{fig:5}
\end{figure*}

\section{discussion} \label{sec:disc}
\begin{figure*}
	\centering
	\includegraphics[width=0.9\textwidth]{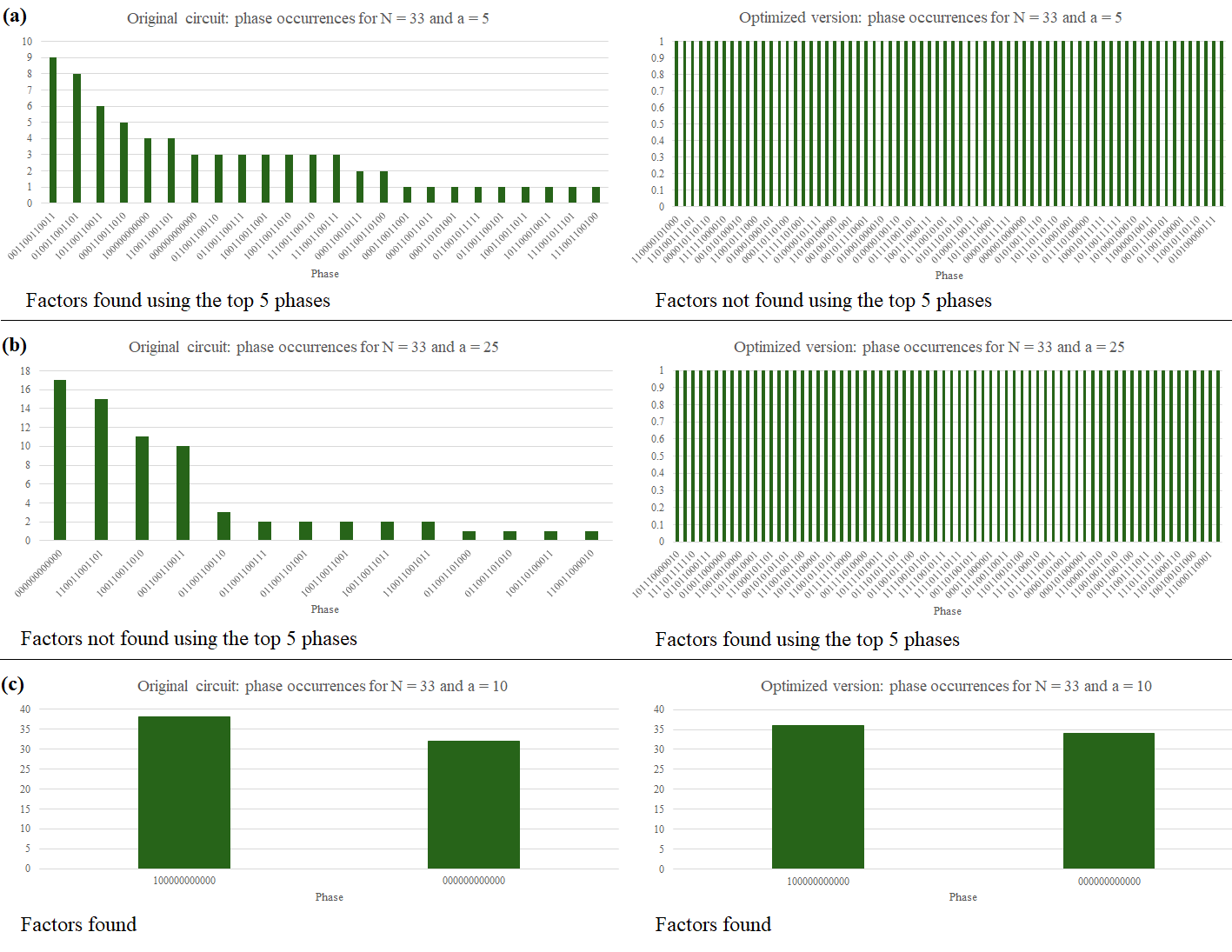}
	\caption{\textbf{Phase histograms of specific couples (N, a) for the original circuit and the optimized version.} Phases were sorted according to their occurrences before plotting. \textbf{(a)} illustrates phases found by the two circuits when $N = 33$ and $a = 5$; the first 5 phases allow factorization only in the original circuit. Differently, histograms represented in \textbf{(b)} consider $N = 33$ and $a = 25$, and entail successful factorization only for the optimized version. Finally, \textbf{(c)} describes very similar phases when $N = 33$ and $a =10$; factors are found by both versions of the circuit.}
	\label{fig:6}
\end{figure*}

Considering the results illustrated in Figure \ref{fig:3}, the success probability of the optimized version is high and similar to the performance of the original circuit, with fewer combinations of $(N, a)$ with $0$ success probability. This is particularly interesting because the approximation introduced on the estimated phase allows to succeed in the  factorization even for values of $a$ that failed in the original circuit.\\\\
The two heat maps show that the occurrences of $100\%$ success probability in the optimized circuit are lower, compared to the original approach. However, this success probability is higher on average and more stable; that is, the success probability of the reduced circuit has a significantly lower standard deviation, compared to the performance of the original circuit. This trend is particularly clear examining Figure \ref{fig:5}, where the success probability is plotted for $N=57$, considering all the possible value for $a$. This performance index significantly oscillates in the original circuit, with many values of $a$ that entail $0$ success probability. Conversely, the optimized circuit has a more stable performance. Indeed, the average success probability for $N=57$ is $68\%$ in the original circuit, while it is $78\%$ for the optimized version.\\\\
The proposed approach outperforms the original circuit in some situations. It is worth noting that for two specific values of $N$, $15$ and $51$, the optimized circuit performs equally to the original one. These two values are product of Fermat numbers ($3, 5, 17$); therefore, this behaviour could be further investigated in the future to determine any potential correlation with the properties of these numbers.\\\\
In addition to the observations above, for each value of $N$, there are some $a$ that entail $100\%$ success probability in the optimized version. Most of them correspond to finding identical phases, compared to the original version. This is further described in Figure \ref{fig:6} which examines $N=33$. Phases found for $a=5$, $a=25$ and $a=10$ are illustrated, considering the original circuit and the optimized version. When $a=5$, the original circuit outperforms the proposed version, while $a=25$ produces the opposite behaviour. Finally, $a=10$ produces $100\%$ success probability for both circuits and it is clear that the same phases are found.\\

\noindent
As it is shown in Figure \ref{fig:heatmap_qpu}, success probability is lower when executing the reduced circuit on the IBM QPU $ibmq\_16\_melbourne$. This behavior matches our expectations, since noise in quantum hardware may introduce errors. However, despite the reduced success probability, factorization was possible on QPU as well. \\\\
Beyond the study on the quality of results, we evaluated the execution time of the optimized circuit on the IBM QPU $ibmq\_16\_melbourne$. These outcomes were compared to the performance of a brute force algorithm on GPU. More specifically, we implemented the trial division method where the target value $N$ is divided by all possible numbers until one factor is found. All the odd numbers between 2 and $\sqrt{N}$ can be considered for this purpose. Tests have been performed on Google Colaboratory and the final results are illustrated in Table \ref{table:1}. We acknowledge that there are faster algorithms for prime factorization, such as the Number Field Sieve \cite{buhler_factoring_1993, crandall_prime_2005}; however, the goal was evaluating our quantum circuit against a simple classical method.

\begin{table}[h!]
\centering
\begin{tabular}{||C{1cm} | C{3cm} | C{3cm}||}
 \hline 
 N & QPU (ms) & GPU (ms)\\ [0.5ex] 
 \hline\hline
 15 & 1595600 & 0.016791 \\ 
 \hline
 21 & 5443000 & 0.017274 \\
 \hline
 33 & 3613700 & 0.015286 \\
 \hline
 57 & 3975600 & 0.017027 \\[0.5ex] 
 \hline
\end{tabular}
\caption{\textbf{Execution time comparison.} The optimized circuit was tested on IBM QPU $ibmq\_16\_melbourne$, while a trial division method was executed on GPU to factorize $15, 21, 33$ and $57$.}
\label{table:1}
\end{table}

\noindent
According to Table \ref{table:1}, the brute force algorithm outperforms the quantum circuit for these small numbers. Nevertheless, we expect QPU running time to decrease with hardware improvements that should be available soon, especially in terms of fast interactions between classical and quantum computing. Finally, the exponential growth in execution times of brute force methods is not observed, considering small numbers only.

\section{Conclusions} \label{sec:concl}
Integer factorization is a mathematical hard problem that represents the core of the RSA cryptosystem. While classical approaches require exponential complexity, Shor's algorithm offers a polynomial time solution to this specific problem. However, both the number of qubits and gates applied, required by the quantum circuit, detrimentally affect the ability to run this circuit on the currently available Quantum hardware. Therefore, we proposed an approximated version of Shor's algorithm to tackle both these problems. \\\\
The introduced approach allows to factor numbers grater than $15$ using a general circuit. Regarding the possible values of $N$, the upper bound depends on the limitations of the considered hardware, in terms of qubits and tolerated depth. As a consequence, we investigated how the optimized circuit scales, when increasing the number to be factored. More specifically, we translated our circuit in an equivalent one that can be executed on a general-purpose quantum hardware with a very limited set of gates ('cx', 'id', 'rz', 'sx', 'x') and full connectivity. Given $n = \lfloor log_{2}(N)\rfloor + 1$, the circuit's depth can be approximated as $10n^{2}+20n+5$. This estimate is a lower bound, since QPUs with different connectivity might introduce more gates to obtain the equivalent circuit. For example, the circuit's depth to factor a key for RSA-1024 would be about $10^{8}$; this is approximately three times the depth experimentally supported by the IBM QPU used to test the circuit. The current quantum hardware suffers from limited connectivity, together with fidelity issues \cite{proctor_measuring_2020}. However, according to IBM's roadmap, QPUs will address these problems in the near future. This might help to further improve the behaviour of the proposed approach. \\\\
With our approach, we were able to factor larger numbers than current state of the art and it was often possible to factor the given number with only one iteration of the algorithm. The approximated circuit may affect the accuracy of the phase estimation, but it does not prevent factorization. For small enough $N$, in some cases, approximated phases allow to factor with values of $a$ that do not work on the original circuit. Therefore, the possibility to execute a reduced version of Shor's algorithm on near term quantum hardware clearly outweighs the introduced approximation errors. \\\\

\section*{Acknowledgments}
\begin{acknowledgments}
This research work was supported and funded by Generali Italia S.p.A. We thank them for fruitful discussions and useful insights.
\end{acknowledgments}

%%% Bibliography %%%%
%%%%%%%%%%%%%
\bibliography{main}{}
\bibliographystyle{ieeetr}

\balancecolsandclearpage

\end{document}